\documentclass[letter,twocolumn]{jpsj3arxiv}

\usepackage{amsmath,amssymb,amsfonts}
\usepackage{graphicx}
\usepackage{bm}

\newcommand{\td}{\mathrm{d}}

\title{Current-Driven Motion of Magnetic Domain Wall with Many Bloch Lines} 

\author{\name{Junichi \surname{Iwasaki}}$^1$\thanks{iwasaki@appi.t.u-tokyo.ac.jp} and 
	\name{Naoto \surname{Nagaosa}}$^{1,2}$\thanks{nagaosa@ap.t.u-tokyo.ac.jp}}
\inst{$^1$Department of Applied Physics, University of Tokyo, Tokyo 113-8656, Japan \\
$^2$RIKEN Center for Emergent Matter Science (CEMS),Wako, Saitama 351-0198, Japan} %\\

\abst{
The current-driven motion of a domain wall (DW) in a ferromagnet with many Bloch lines (BLs) via the spin transfer torque is studied theoretically.
It is found that the motion of BLs changes the current-velocity ($j$-$v$) characteristic dramatically.
Especially, the critical current density to overcome the pinning force is reduced by the factor of the Gilbert damping coefficient $\alpha$ even compared with that of a skyrmion.
This is in sharp contrast to the case of magnetic field driven motion, where the existence of BLs reduces the mobility of the DW.
}

\begin{document}
\maketitle

Domain walls (DWs) and bubbles~\cite{Hubert98magnetic,Malozemoff79magnetic} are the spin textures in ferromagnets which have been studied intensively over decades from the viewpoints of both fundamental physics and applications.
The memory functions of these objects are one of the main focus during 70's, but their manipulation in terms of the magnetic field faced the difficulty associated with the pinning which hinders their motion.
The new aspect introduced recently is the current-driven motion of the spin textures~\cite{Slonczewski96current,Berger96emission}.
The flow of the conduction electron spins, which follow the direction of the background localized spin moments, moves the spin texture due to the conservation of the angular momentum.
This effect, so called the spin transfer torque, is shown to be effective to manipulate the DWs and bubbles compared with the magnetic field.
Magnetic skyrmion~\cite{Mulbauer09skyrmion,Yu10real} is especially an interesting object, which is a swirling spin texture acting as an emergent particle protected by the topological invariant, i.e., the skyrmion number $N_\mathrm{sk}$, defined by
\begin{equation}
N_\mathrm{sk} = \frac{1}{4 \pi} \int \td^2 r \ \bm{n}(\bm{r}) \cdot \left(
\frac{\partial \bm{n}(\bm{r})}{\partial x}
\times
\frac{\partial \bm{n}(\bm{r})}{\partial y} \right)
\end{equation}
with $\bm{n}(\bm{r})$ being the unit vector representing the direction of the spin as a function of the two-dimensional spatial coordinates $\bm{r}$.
This is the integral of the solid angle subtended by $\bm{n}$, and counts how many times the unit sphere is wrapped.
The solid angle and skyrmion number $N_\mathrm{sk}$ also play essential role when one derives the equation of motion for the center of mass of the spin texture, i.e., the gyro-motion is induced by $N_\mathrm{sk}$ in the Thiele equation, where the rigid body motion is assumed~\cite{Thiele73steady,Everschor12rotating}.

\begin{figure}
\begin{center}
\includegraphics[width=\hsize]{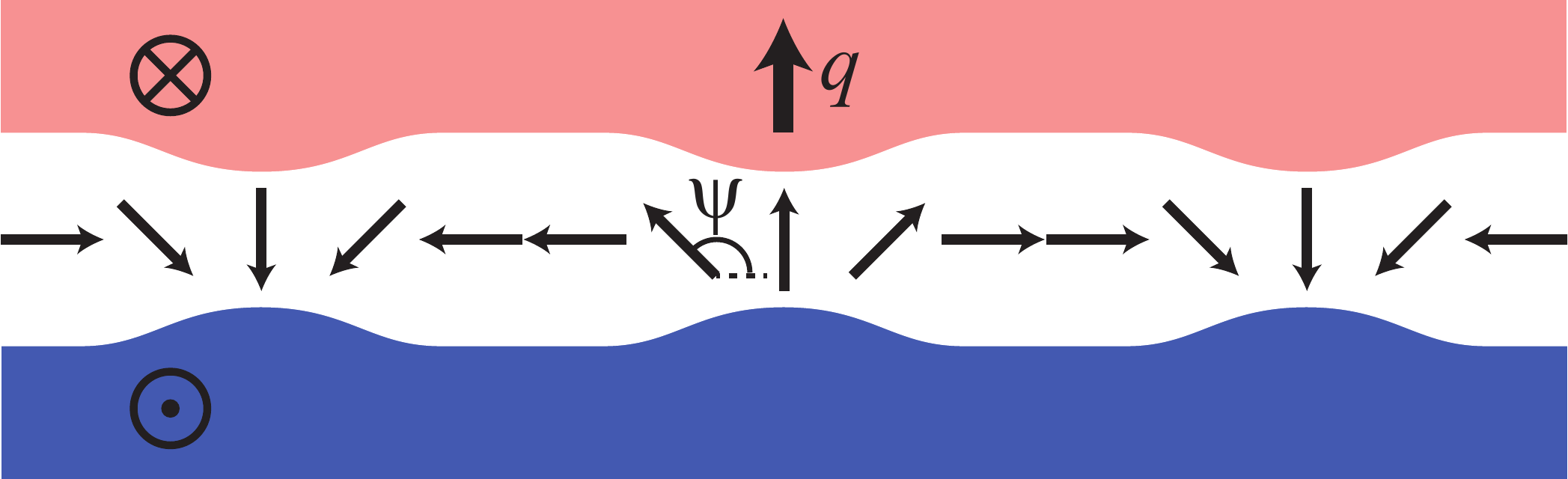}
\end{center}
\caption{
Schematic magnetization distribution of DW with many Bloch lines.
}
\label{fig:schematic}
\end{figure}

Beyond the Thiele equation~\cite{Thiele73steady}, one can derive the equation of motion of a DW in terms of two variables, i.e., the wall-normal displacement $q(t,\zeta,\eta)$ and the wall-magnetization orientation angle $\psi(t,\zeta,\eta)$ (see Fig.~\ref{fig:schematic}) where $\zeta$ and $\eta$ are general coordinates specifying the point on the DW~\cite{Slonczewski74theory}:
\begin{align}
\frac{\delta \sigma}{\delta \psi}
=  2 M \gamma^{-1} \left[ \dot{q} - \alpha \Delta \dot{\psi}  - v^\mathrm{s}_\perp - \beta \Delta v^\mathrm{s}_\parallel (\partial_\parallel \psi ) \right], 
\label{eq:EOM1}
\end{align}
\begin{align}
\frac{\delta \sigma}{\delta q} = - 2 M \gamma^{-1} \left[ \dot{\psi} + \alpha \Delta^{-1} \dot{q} + v^\mathrm{s}_\parallel(\partial_\parallel \psi) - \beta \Delta^{-1} v^\mathrm{s}_\perp \right], 
\label{eq:EOM2}
\end{align}
Here, $\dot{}$ means the time-derivative.
$\parallel$ and $\perp$ indicate the components parallel and perpendicular to the DW respectively.
$M$ is the magnetization, $\gamma$ is the gyro-magnetic ratio, and $\sigma$, $\Delta$ are the energy per area and thickness of the DW.
$v^\mathrm{s}$ is the velocity of the conduction electrons, which produces the spin transfer torque.
$\alpha$ is the Gilbert damping constant, and $\beta$ represents the non-adiabatic effect.
These equations indicate that $q$ and $\psi$ are canonical conjugate to each other.
This is understood by the fact that the generator of the spin rotation normal to the DW, which is proportional to $\sin \psi$ in Fig.~\ref{fig:schematic}, drives the shift of $q$.
(Note that $\psi$ is measured from the fixed direction in the laboratory coordinates.)

In order to reduce the magnetostatic energy, the spins in the DW tend to align parallel to the DW, i.e., Bloch wall.
When the DW is straight, this structure is coplanar and has no solid angle.
From the viewpoint of eqs.~\eqref{eq:EOM1} and \eqref{eq:EOM2}, the angle $\psi$ is fixed around the minimum, and slightly canted when the motion of $q$ occurs, i.e., $\dot{\psi}=0$.
However, it often happens that the Bloch lines (BLs) are introduced into the DW as shown schematically in Fig.~\ref{fig:schematic}.
The angle $\psi$ rotates along the DW and the N\'eel wall is locally introduced.
It is noted here that the solid angle becomes finite in the presence of the BLs.
Also with many BLs in the DW, the translation of BLs activates the motion of the angle $\psi$, i.e., $\dot{\psi} \ne 0$, which leads to the dramatic change in the dynamics.

In the following, we focus on the straight DW which extends along $x$-direction and is uniform in $z$-direction.
Thus, the general coordinates here are $(\zeta,\eta)=(x,z)$.
$q(t,x,z)$ is independent of the coordinates $q(t,x,z)=q(t)$, and the functional derivative $\delta \sigma/\delta q$ in eq.~\eqref{eq:EOM2} becomes the partial derivative $\partial \sigma/\partial q$.
In the absence of BLs, we set $\psi(t,x,z)=\psi(t)$, and $\delta \sigma/\delta \psi$ in eq.~\eqref{eq:EOM1} also becomes $\partial \sigma/\partial \psi$.
Then the equation of motion in the absence of BL is
\begin{align}
\frac{\partial \sigma}{\partial \psi}
=  2 M \gamma^{-1} \left[ \dot{q} - \alpha \Delta \dot{\psi}  - v^\mathrm{s}_\perp \right], 
\label{eq:EOM1woBL}
\end{align}
\begin{align}
\frac{\partial \sigma}{\partial q} = - 2 M \gamma^{-1} \left[ \dot{\psi} + \alpha \Delta^{-1} \dot{q} - \beta \Delta^{-1} v^\mathrm{s}_\perp \right], 
\label{eq:EOM2woBL}
\end{align}
With many BLs, the sliding motion of Bloch lines along DW, which activates $\dot{\psi}$, does not change the wall energy, i.e., $\delta \sigma / \delta \psi$ in eq.~\eqref{eq:EOM1} vanishes~\cite{Malozemoff79magnetic}.
Here, for simplicity, we consider the periodic BL array with the uniform twist $\psi(t,x,z)=(x-p(t))/\tilde{\Delta}$ where $\tilde{\Delta}$ is the distance between BLs, which leads to
\begin{align}
0 =  2 M \gamma^{-1} \left[ \dot{q} + \alpha \Delta \tilde{\Delta}^{-1} \dot{p}  - v^\mathrm{s}_\perp - \beta \Delta \tilde{\Delta}^{-1} v^\mathrm{s}_\parallel \right], 
\label{eq:EOM1wBL}
\end{align}
\begin{align}
\frac{\partial \sigma}{\partial q} = - 2 M \gamma^{-1} \left[ - \tilde{\Delta}^{-1} \dot{p} + \alpha \Delta^{-1} \dot{q} + \tilde{\Delta}^{-1} v^\mathrm{s}_\parallel - \beta \Delta^{-1} v^\mathrm{s}_\perp \right], 
\label{eq:EOM2wBL}
\end{align}

First, let us discuss the magnetic field driven motion without current.
The effect of the external magnetic field $H^\mathrm{ext}$ is described by the force $\partial \sigma / \partial q = -2MH^\mathrm{ext}$ in eqs.~\eqref{eq:EOM2woBL} and \eqref{eq:EOM2wBL}.
$v^\mathrm{s}_\parallel$ and $v^\mathrm{s}_\perp$ are set to be zero.
In the absence of BL, as mentioned above, the phase $\psi$ is static $\dot{\psi}=0$ with the slight tilt of the spin from the easy-plane, and one obtains from eq.~\eqref{eq:EOM2woBL}
\begin{align}
\dot{q} = \frac{\Delta \gamma H^\mathrm{ext}}{\alpha}.
\label{eq:qmag}
\end{align}
This is a natural result, i.e., the mobility is inversely proportional to the Gilbert damping $\alpha$.
$\psi$ is determined by eq.~\eqref{eq:EOM1woBL} with this value of the velocity $\dot{q}$.

In the presence of many BLs, eqs.~\eqref{eq:EOM1wBL} and \eqref{eq:EOM2wBL} give the velocities of DW and BL sliding driven by the magnetic field as
\begin{align}
\dot{q} = \frac{\alpha}{1 + \alpha^2} \Delta \gamma H^\mathrm{ext},
\label{eq:qmagBL}
\end{align}
\begin{align}
\dot{p} = - \frac{1}{1 + \alpha^2} \tilde{\Delta} \gamma H^\mathrm{ext}.
\label{eq:pmagBL}
\end{align}
Comparing eqs.~\eqref{eq:qmag} and \eqref{eq:qmagBL}, the mobility of the DW is reduced by the factor of $\alpha^2$ since $\alpha$ is usually much smaller than unity.
We also note that the velocity of the BL sliding $\dot{p}$ is larger than that of the wall $\dot{q}$ by the factor of $\alpha$.
Physically, this means that the effect of the external magnetic field $H^\mathrm{ext}$ mostly contributes to the rapid motion of the BLs along the DW rather than the motion of the DW itself.
These results have been already reported in refs.~\cite{Malozemoff79magnetic,Slonczewski74theory,Malozemoff72effect}.
 
Now let us turn to the motion induced by the current $v^\mathrm{s}$.
In the absence of BL, again we put $\dot{\psi}=0$ in eqs.~\eqref{eq:EOM1woBL} and \eqref{eq:EOM2woBL}.
Assuming that there is no pinning force or external magnetic field, i.e., $\partial \sigma / \partial q=0$, one obtains from eq.~\eqref{eq:EOM2woBL}
\begin{align}
\dot{q} = \frac{\beta}{\alpha} v^\mathrm{s}_\perp,
\label{eq:current}
\end{align}
and eq.~\eqref{eq:EOM1woBL} determines the equilibrium value of $\psi$.
When the pinning force $\partial \sigma / \partial q =F^\mathrm{pin}$ is finite, there appears a threshold current density $\left( v^\mathrm{s}_\perp \right)_\mathrm{c}$ which is determined by putting $\dot{q} = 0$ in eq.~\eqref{eq:EOM2woBL} as
\begin{align}
\left( v^\mathrm{s}_\perp \right)_\mathrm{c} = \frac{\gamma \Delta}{2 M \beta} F^\mathrm{pin},
\label{eq:critical}
\end{align}
which is inversely proportional to $\beta$~\cite{Tatara06threshold}.
Since eq.~\eqref{eq:current} is independent of $v^\mathrm{s}_\parallel$, the threshold current density $\left( v^\mathrm{s}_\parallel \right)_\mathrm{c}$ is $\left( v^\mathrm{s}_\parallel \right)_\mathrm{c} = \infty$.

In the presence of the many BLs, on the other hand, eqs.~\eqref{eq:EOM1wBL} and \eqref{eq:EOM2wBL} give
\begin{align}
\frac{\partial \sigma}{\partial q} = - 2M \gamma^{-1} 
&\left[  \frac{1 + \alpha^2}{\alpha} \Delta^{-1} \dot{q} \right. \notag \\
&\left. - \frac{1 + \alpha \beta}{\alpha} \Delta^{-1} v^\mathrm{s}_\perp
-\frac{\beta-\alpha}{\alpha} \tilde{\Delta}^{-1} v^\mathrm{s}_\parallel \right],
\label{eq:cBL}
\end{align}
which is the main result of this paper.
From eq.~\eqref{eq:cBL}, the current-velocity characteristic in the absence of both the pinning and the external field ($\partial \sigma / \partial q$=0) is
\begin{align}
\dot{q} &= \frac{1 + \alpha \beta}{1+ \alpha^2} v^\mathrm{s}_\perp - \frac{\beta-\alpha}{1+ \alpha^2} \Delta \tilde{\Delta}^{-1} v^\mathrm{s}_\parallel \notag \\
&\simeq v^\mathrm{s}_\perp + (\beta-\alpha) \Delta \tilde{\Delta}^{-1} v^\mathrm{s}_\parallel,
\label{eq:univ}
\end{align}
where the fact $\alpha, \beta \ll 1$ is used in the last step.
If we neglect the term coming from $v^\mathrm{s}_\parallel$, the current-velocity relation becomes almost independent of $\alpha$ and $\beta$ in sharp contrast to eq.~\eqref{eq:current}.
This is similar to the universal current-velocity relation in the case of skyrmion~\cite{Iwasaki13universal}, where the solid angle is finite and also the transverse motion to the current occurs.
Note that $v^\mathrm{s}_\parallel$ slightly contributes to the motion when $\alpha \ne \beta$, while it does not in the absence of BL.
Even more dramatic is the critical current density in the presence of the pinning ($\partial \sigma / \partial q = F^\mathrm{pin}$).
When we apply only the current perpendicular to the DW, i.e., $v^\mathrm{s}_\parallel = 0$, putting $\dot{q}=0$ in eq.~\eqref{eq:cBL} determines the threshold current density as
\begin{align}
\left( v^\mathrm{s}_\perp \right)_\mathrm{c} = \frac{\gamma \Delta}{2 M}
\frac{\alpha}{1 + \alpha \beta} F^\mathrm{pin},
\label{eq:critical2}
\end{align}
which is much reduced compared with eq.~\eqref{eq:critical} by the factor of $\frac{\alpha \beta}{1 + \alpha \beta} \ll 1$.
Note that $\left( v^\mathrm{s}_\perp \right)_\mathrm{c}$ in eq.~\eqref{eq:critical2} is even smaller than the case of skyrmion~\cite{Iwasaki13universal} by the factor of $\alpha$.
Similarly, the critical current density of the motion driven by $v^\mathrm{s}_\parallel$ is given by
\begin{align}
\left( v^\mathrm{s}_\parallel \right)_\mathrm{c} = \frac{\gamma \tilde{\Delta}}{2 M}
\frac{\alpha}{| \beta- \alpha |} F^\mathrm{pin},
\label{eq:critical3}
\end{align}
which can also be smaller than eq.~\eqref{eq:critical}.

\begin{figure}
\begin{center}
\includegraphics[width=\hsize]{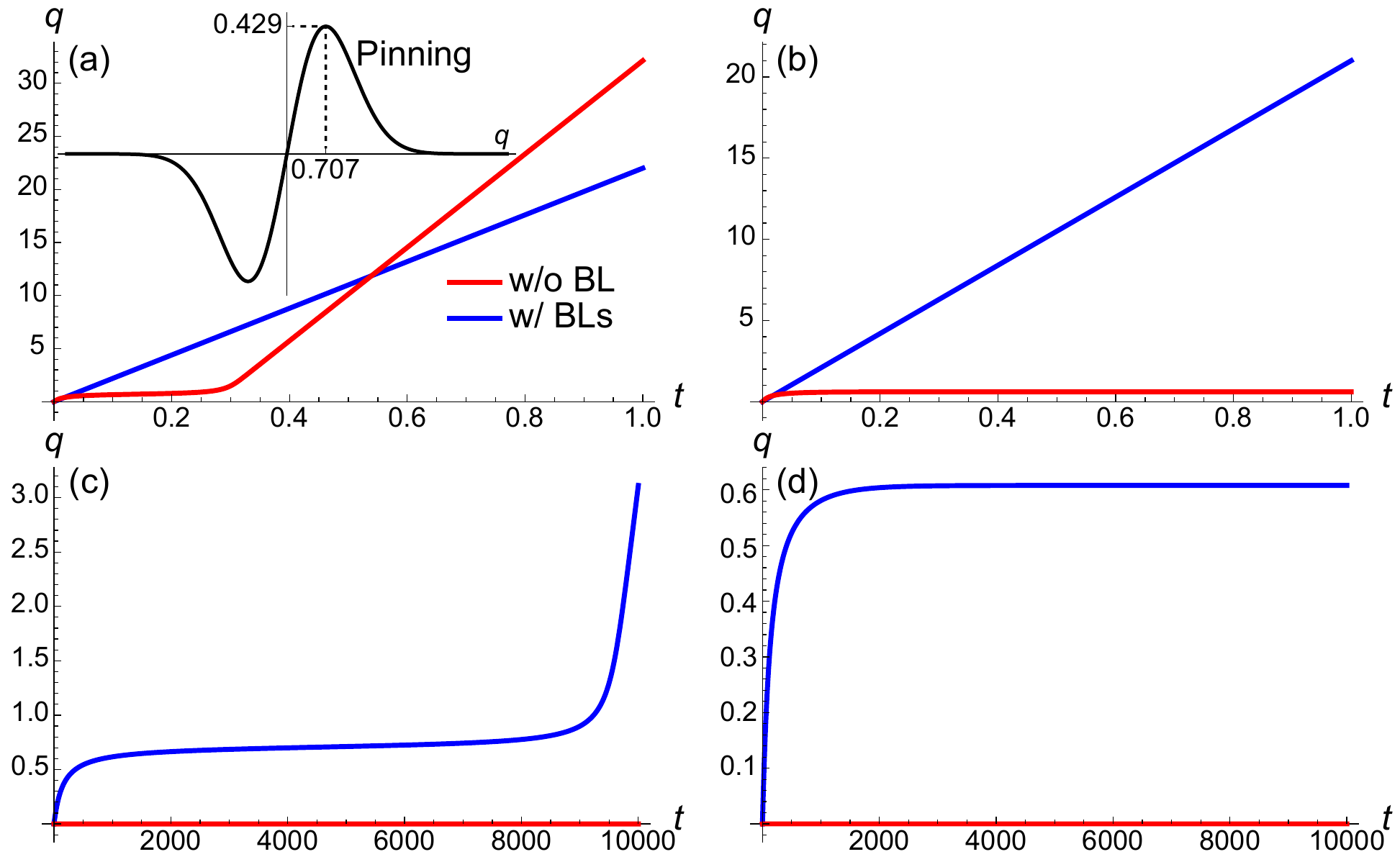}
\end{center}
\caption{
The wall displacement $q$ as a fucntion of $t$ for the DWs without BL and with BLs.
(a)
$v^\mathrm{s}_\perp=22.0$.
The inset shows the pinning force $F^\mathrm{pin}$.
(b)
$v^\mathrm{s}_\perp=21.0$.
(c)
$v^\mathrm{s}_\perp=0.0043$.
(d)
$v^\mathrm{s}_\perp=0.0042$.
}
\label{fig:num_sol}
\end{figure}

Next we look at the numerical solutions of $q(t)$ driven by the current $v^\mathrm{s}_\perp$ perpendicular to the wall under the pinning force.
We assume the following pinning force: $(\gamma \Delta/2M)F^\mathrm{pin}(q) = v^\ast (q/\Delta) \exp\left[ -(q/\Delta)^2 \right]$ (see the inset of Fig.~\ref{fig:num_sol}(a)).
We employ the unit of $\Delta = v^\ast = 1$ and the parameters $(\alpha, \beta)$ are fixed at $(\alpha, \beta)=(0.01,0.02)$.
Here, we compare two DWs without BL and with BLs.
The maximum value of the pinning force $(\gamma \Delta/2M) F^\mathrm{pin}_\mathrm{max} = 0.429$ determines the threshold current density $\left( v^\mathrm{s}_\perp \right)_\mathrm{c}$ as $\left( v^\mathrm{s}_\perp \right)_\mathrm{c} = 21.4$ and $\left( v^\mathrm{s}_\perp \right)_\mathrm{c} = 0.00429$ in the absence of BL and in the presence of many BLs, respectively.
In Fig.~\ref{fig:num_sol}(a), both DWs overcome the pinning at the current density $v^\mathrm{s}_\perp=22.0$, although the velocity of the DW without BL is suppressed in the pinning potential.
At the current density $v^\mathrm{s}_\perp=21.0$ below the threshold value in the absence of BL, the DW without BL is pinned, while that with BLs still moves easily (Fig.~\ref{fig:num_sol}(b)).
The velocity suppression in the presence of BLs is observed at much smaller current density $v^\mathrm{s}_\perp=0.0043$ (Fig.~\ref{fig:num_sol}(c)), and finally it stops at $v^\mathrm{s}_\perp=0.0042$ (Fig.~\ref{fig:num_sol}(d)).

All the discussion above relies on the assumption that the wall is straight and $\psi$ rotates uniformly.
When the bending of the DW and non-uniform distribution of BLs are taken into account, the average velocity and the threshold current density take the values between two cases without BL and with many BLs.
The situation changes when the DW forms closed loop, i.e., the domain forms a bubble.
The bubble with many BLs and large $|N_\mathrm{sk}|$ is called {\it hard bubble} because the repulsive interaction between the BLs makes it hard to collapse the bubble~\cite{Malozemoff79magnetic}.
At the beginning of the motion, the BLs move along the DW, which results in the tiny critical current.
In the steady state, however, the BLs accumulate in one side of the bubble~\cite{Vella-Coleiro72dynamic,Thiele73dynamic}.
Then, the configuration of the BLs is static and the Thiele equation is justified as long as the force is slowly varying within the size of the bubble.
The critical current density $\left( v^\mathrm{s} \right)_\mathrm{c}$ is given by $\left( v^\mathrm{s} \right)_\mathrm{c} \propto F^\mathrm{pin} / N_\mathrm{sk}$ ($N_\mathrm{sk}$ ($\gg 1$): the skyrmion number of the hard bubble), and is reduced by the factor of $N_\mathrm{sk}$ compared with the skyrmion with $N_\mathrm{sk} = \pm 1$.

In conclusion, we have studied the current-induced dynamics of the DW with many BLs.
The finite $\dot{\psi}$ in the steady motion activated by BLs sliding drastically changes the dynamics, which has already been reported in the field-driven case.
In contrast to the field-driven case, where the mobility is suppressed by introducing BLs, that in the current-driven motion is not necessarily suppressed.
Instead, the current-velocity relation shows universal behavior independent of the damping strength $\alpha$ and non-adiabaticity $\beta$.
Furthermore, the threshold current density in the presence of impurities is tiny even compared with that of skyrmion motion by the factor of $\alpha$.
These findings will stimulate the development of the racetrack memory based on the DW with many BLs.

\acknowledgments{
We thank W. Koshibae for useful discussion.
This work is supported by Grant-in-Aids for Scientific Research (S) (No. 24224009) from the Ministry of Education, Culture, Sports, Science and Technology of Japan.
J. I. was supported by Grant-in-Aids for JSPS Fellows (No. 2610547).
}

%\bibliography{test}

\end{document}